\documentclass[journal,twoside,web]{ieeecolor}
\usepackage{lcsys}
\usepackage{cite}
\usepackage{amsmath,amssymb,amsfonts}
\usepackage{algorithmic}
\usepackage{graphicx}
\usepackage{textcomp}
\usepackage{url}
\usepackage{hyperref}
\usepackage{cancel}

\def\BibTeX{{\rm B\kern-.05em{\sc i\kern-.025em b}\kern-.08em
    T\kern-.1667em\lower.7ex\hbox{E}\kern-.125emX}}
\begin{document}
\title{Exponentially Stable MRAC of MIMO Switched Systems with Matched Uncertainty and Completely Unknown Control Matrix}
\author{A. Glushchenko, \IEEEmembership{Member, IEEE}, and K. Lastochkin
\thanks{This research was financially supported in part by Grants Council of the President of the Russian Federation (project MD-1787.2022.4). }
\thanks{Anton Glushchenko is with Laboratory 7 of V.A. Trapeznikov Institute of Control Sciences of RAS, Moscow, Russia (e-mail: aiglush@ipu.ru). }
\thanks{Konstantin Lastochkin is with Laboratory 7 of V.A. Trapeznikov Institute of Control Sciences of RAS, Moscow, Russia (e-mail: lastconst@ipu.ru).}}

\maketitle

\thispagestyle{empty} 

\begin{abstract}
In this paper an attempt is made to extend the concept of the exponentially stable adaptive control to one class of multi-input-multi-output (MIMO) plants with matched nonlinearity and unknown piecewise constant parameters. Within the intervals between two consecutive parameter switches, the proposed adaptive control system ensures: 1) exponential convergence to zero of the parameter and reference model tracking errors, 2) the monotonicity of the control law adjustable parameters. Both properties are guaranteed in case the regressor is finitely exciting somewhere inside each of such intervals. Compared to the existing methods, the proposed one is applicable to systems with unknown switching signal function and completely unknown control matrix. The theoretical results are supported by the numerical experiments.
\end{abstract}


\section{Introduction}
\label{sec:introduction}
\IEEEPARstart{M}{ethods} of the model reference adaptive control (MRAC) theory are aimed at solving control problems under the conditions of significant parameter uncertainty. Such classical problems are the adjustment of the state- or output-feedback control law parameters under the influence of matched/unmatched, structured/unstructured disturbances and the uncertainty of the state $A \in {\mathbb{R}^{n \times n}}$ and control  $B \in {\mathbb{R}^{n \times m}}$ matrices \cite{b1, b2}. Nowadays, the effective synthesis procedures have been proposed that guarantee the asymptotic convergence of the plant trajectories to reference ones in all stated cases. Efficient robust modifications are developed to ensure boundedness of all signals and certain trajectory tracking quality for a system being affected by external disturbances \cite{b3}, unmodeled dynamics \cite{b4}, and delays \cite{b5}.

However, even under ideal conditions, the classical approaches to adaptive systems synthesis: \emph{i}) guarantee both the exponential stability of a closed-loop system and the convergence of controller adjustable parameters to their true values only when the strict requirement of the regressor persistent excitation (PE) is met \cite{b6}, \emph{ii}) require {\it a priori} information about the control matrix \cite{b1,b2,b7}. The mentioned disadvantages significantly limit the applicability of adaptive control systems to real-world problems, so that considerable attention of control science community has been attracted to overcome them in recent years.

In \cite{b8,b9,b10}, various methods have been developed to relax the PE requirement to ensure parameter error exponential convergence. In \cite{b11,b12,b13,b14}, approaches are proposed that simultaneously relax the PE condition and reduce the amount of the required {\it a priori} data on the control matrix. The disadvantages of \cite{b8,b9,b10} are that the adaptive laws are insensitive to switches of the plant unknown parameters \cite{b11}, and only the norm of the unknown parameters identification error is monotonic. In turn, \cite{b11} requires the matrix $A$ to be known, the method from \cite{b12} demands the known low bound of $det \left\{ {{B^{\rm{T}}}B} \right\}$, whereas \cite{b13} – of $det \left\{ {{{\left( {{B^{\rm{\dagger }}}{B_{ref}}} \right)}^{ - 1}}} \right\}$, and ${\rm{sgn}}\left\{ {det \left\{ {{B^{\rm{\dagger }}}{B_{ref}}} \right\}} \right\}$ is needed in \cite{b14}.

To overcome the shortcomings of the solutions \cite{b8,b9,b10,b11,b12,b13,b14}, a new concept of exponentially stable adaptive control has been proposed recently \cite{b15}. The adaptive law is derived in two steps. In the first one, considering the plant equations, measurable regressions ${z_A}\left( t \right) = \Delta \left( t \right)A{\rm{,\;}}{z_B}\left( t \right) = \Delta \left( t \right)B$ with a scalar regressor $\Delta \in \mathbb{R}$ with respect to the unknown matrices $A{\rm{,\;}}B$ are parameterized using stable minimum-phase filters and the dynamic regressor extension and mixing technique \cite{b16}. At the second step, such regressions are substituted into the {\it a priori} known equations to calculate the control law ideal parameters (e.g. $A + B{K_x} = {A_{ref}}{\rm{,\;}}B{K_r} = {B_{ref}}$) to obtain measurable regression equations (e.g. ${z_{{K_x}}}\!\left(t \right)\!\! =\!\! \Omega \!\left( t \right)\!{K_x}$, ${z_{{K_r}}}\left( t \right) = \Omega \left( t \right){K_r}$) again with a scalar regressor $\Omega \in \mathbb{R}$, but now with respect to the mentioned ideal parameters of the control law. Then, on the basis of such a regression, the second Lyapunov method and the results of the Kalman-Yakubovich-Popov Lemma are applied to derive the adaptive law to adjust the controller parameters, which does not require any {\it a priori} information about $B$ and ensures the exponential stability of the closed-loop system as well as the monotonic transients of each controller unknown parameter estimate. This approach significantly differs from both conventional indirect and direct MRAC and is the closest ideologically to self-tuning control systems, also some parallels can be found with the Monopoli's augmented error scheme \cite{b1, b2}. The concept of the exponentially stable adaptive control has been successfully applied to several state-feedback problems of: 1) MRAC-based control of MIMO \cite{b17} and SISO systems \cite{b15}, 2) the direct adaptive pole placement control \cite{b18}, 3) the adaptive optimal linear quadratic regulation \cite{b19}.

The main general drawback of \cite{b15, b17,b18,b19}, as well as the whole concept of the exponentially stable adaptive control, is the applicability of the proposed adaptive control systems only to linear plants with time-invariant unknown parameters, as well as the requirement to meet the condition of regressor finite excitation to ensure the asymptotic stability of the reference model states tracking error.

The aim of this study is to extend the concept of the exponentially stable adaptive control \cite{b15} to one class of MIMO systems with matched nonlinearity and piecewise constant unknown parameters. The main contribution of the proposed solution with respect to \cite{b15, b17,b18,b19} is threefold:
\begin{enumerate}
    \item [\textbf{C1}] an adaptive control approach for MIMO systems with matched nonlinearity and piecewise constant unknown parameters, which does not require {\it a priori} information about the matrix $B$, is proposed;
    \item [\textbf{C2}] the parameter and reference model tracking errors converge exponentially to zero within the time intervals between consecutive switches of the unknown parameters;
    \item [\textbf{C3}] the monotonicity of the control law adjustable parameters is guaranteed within the time ranges stated in (C2).
\end{enumerate}

To achieve such properties, it is proposed to augment the parameterization used in \cite{b15, b17,b18,b19} to obtain ${z_{{K_x}}}\left( t \right){\rm{,\;}}{z_{{K_r}}}\left( t \right)$ by a recently proposed algorithm to detect time instants of switches of the linear regression equation unknown parameters \cite{b20}, and to set to zero the outputs of all dynamic filters of the parameterization when such switch is detected. Assuming that the regressor is finitely exciting somewhere inside the time interval between two consecutive switches, such an approach allows one to obtain a measurable regression equation with respect to the current values of the unknown parameters of the control law, which in accordance with \cite{b15, b17,b18,b19} is necessary and sufficient for exponential convergence of both tracking and parameter error within the considered time intervals.

Here we would also like to note that filtering with resetting was previously used in the parameterization of the adaptive control problem in \cite{b21}. However, in contrast to this study, the parametrization from \cite{b21} requires one to know the matrix $B$, and the filters are set to a zero state not when the switch of the plant parameters is detected, but when the reference value is changed. Therefore, in comparison with \cite{b21}, the proposed solution ensures an improved transient quality and can be applied to problems with unknown matrix $B$.

The structure of the paper is arranged as follows. The generalized problem statement is presented in Section 2. Section 3 describes the proposed adaptive system for MIMO switched systems with matched uncertainty and unknown control matrix. The simulation results are shown in Section 4. The paper is wrapped-up with conclusion in Section 5.

\subsection{Definitions}
The definition of the regressor finite excitation and the corollary of the Kalman-Yakubovich-Popov lemma are used in the proofs of theorem and propositions.

\textbf{\emph{Definition.}}\emph{ A regressor $\omega \left( t \right)$ is finitely exciting $\omega \left( t \right) \in {\rm{FE}}$ over a time range $\left[ {t_r^ + {\rm{;\;}}{t_e}} \right]{\rm{,}}$ if there exist $t_r^ +  \ge 0$, ${t_e} \ge t_r^ + $ and $\alpha $ such that the following inequality holds:
\begin{equation}\label{eq1}
\int\limits_{t_r^ + }^{{t_e}} {\omega \left( \tau  \right){\omega ^{\rm{T}}}\left( \tau  \right)d} \tau  \ge \alpha I{\rm{,}}
\end{equation}
where $\alpha > 0$ is the excitation level, $I$ is an identity matrix.
}

\textbf{\emph{Corollary 1.}}\emph{ For any matrix $D > 0$, scalar $\mu  > 0$, controllable pair $\left( {A{\rm{,\;}}B} \right)$ with $B \in {\mathbb{R}^{n \times m}}$ and Hurwitz matrix $A \in {\mathbb{R}^{n \times n}}$ there exist a matrix $P = {P^{\rm{T}}} > 0$ \linebreak and $Q \in {\mathbb{R}^{n \times m}},\;K \in {\mathbb{R}^{m \times m}}$, such that:
\begin{equation}\label{eq2}
  \begin{gathered} 
\begin{array}{c}
{A^{\rm{T}}}P + PA =  - Q{Q^{\rm{T}}} - \mu P{\rm{,\;}}PB = QK{\rm{,}}\\
{K^{\rm{T}}}K = D + {D^{\rm{T}}}.
\end{array}
  \end{gathered}
\end{equation}
}

\section{Problem Statement}

A class of linear MIMO switched plants with matched nonlinear uncertainty is considered:
\begin{equation}{{\label{eq3}}}
\begin{array}{c}
\forall t \ge t_0^ + {\rm{,\;}}\dot x\left( t \right) = {\Theta ^{\rm{T}}}\left( t \right)\Phi \left( t \right) = \\
\left\{ \begin{array}{l}
{A_0}x\left( t \right) + {B_0}\left( {u\left( t \right) + \vartheta _0^{\rm{T}}\Psi \left( x \right)} \right){\rm{,\;}}t \in \left[ {t_0^ + {\rm{;\;}}t_1^ + } \right)\\
 \vdots \\
{A_i}x\left( t \right) + {B_i}\left( {u\left( t \right) + \vartheta _i^{\rm{T}}\Psi \left( x \right)} \right){\rm{,\;}}t \in \left[ {t_i^ + {\rm{;\;}}t_{i + 1}^ + } \right)
\end{array} \right.\\
x\left( {t_0^ + } \right) = {x_0}{\rm{,}}\\
\Phi \left( t \right) = {{\begin{bmatrix}
{{x^{\rm{T}}}\left( t \right)}&{{u^{\rm{T}}}\left( t \right)}&{{\Psi ^{\rm{T}}}\left( x \right)}
\end{bmatrix}}^{\rm{T}}}{\rm{,}}\\
{\Theta _i} = {\begin{bmatrix}
{{A_i}}&{{B_i}}&{{B_i}\vartheta _i^{\rm{T}}}
\end{bmatrix}} {\rm{,\;}}i \in \mathbb{N}{\rm{,}}
\end{array}
\end{equation}
where $x\left( t \right) \in {\mathbb{R}^n}$ is a vector of plant states with unknown initial conditions ${x_0}$, $u\left( t \right) \in {\mathbb{R}^m}$ is a control signal, \linebreak $\Psi \left( x \right) \in {\mathbb{R}^p}$ are known basis functions, ${A_i} \in {\mathbb{R}^{n \times n}}$ is an unknown state matrix, ${B_i} \in {\mathbb{R}^{n \times m}}$ is an unknown control matrix, ${\vartheta _i} \in {\mathbb{R}^{p \times m}}$ is a matrix of unknown parameters of the uncertainty, $t_0^ + $ is a known initial time instant. The pair $\left( {{A_i}{\rm{,\;}}{B_i}} \right)$ is controllable, $\forall t > t_0^ + $ the vector $\Phi \left( t \right) \in {\mathbb{R}^{n + m + p}}$ is measurable, and the matrix ${\rm{\;}}{\Theta _i} \in {\mathbb{R}^{\left( {n + m + p} \right) \times n}}$ and the switching time instants $t_i^ + $ are unknown $\forall i > 0$.

The required transient quality for the plant \eqref{eq3} is defined by the following reference model:
\begin{equation}\label{eq4}
\begin{array}{c}
\forall t \ge t_0^ + {\rm{,\;}}{\dot x_{ref}}\left( t \right) = {A_{ref}}{x_{ref}}\left( t \right) + {B_{ref}}r\left( t \right){\rm{,\;}}\\
{x_{ref}}\left( {t_0^ + } \right) = {x_{0ref}}{\rm{,}}
\end{array}
\end{equation}
where ${x_{ref}}\left( t \right) \in {\mathbb{R}^n}$ is a reference model (RM) state vector, ${x_{ref}}\left( {t_0^ + } \right) \in {\mathbb{R}^n}$ is a vector of initial conditions, $r\left( t \right) \in {\mathbb{R}^m}$ is a reference signal, ${A_{ref}} \in {\mathbb{R}^{n \times n}}$ is a Hurwitz state matrix of RM, ${B_{ref}} \in {\mathbb{R}^{n \times m}}$ is a RM control matrix.

Ideal model following conditions are assumed to be met:

\emph{ \textbf{Assumption 1.} There exist $K_i^x \in {\mathbb{R}^{m \times n}}$, $K_i^r \in {\mathbb{R}^{m \times m}}$ such that:}
\begin{equation}\label{eq5}
{A_i} + {B_i}K_i^x = {A_{ref}}{\rm{,\;}}{B_i}K_i^r = {B_{ref}}.
\end{equation}

Considering Assumption 1, the error equation between \eqref{eq3} and \eqref{eq4} is written as:
\begin{equation}\label{eq6}
\begin{array}{c}
{{\dot e}_{ref}}\left( t \right) = {A_{ref}}{e_{ref}}\left( t \right) + {B_i}\left( {u\left( t \right) + \vartheta _i^{\rm{T}}\Psi \left( x \right)} \right) -\\
-\left( {{A_{ref}} - {A_i}} \right)x\left( t \right) - {B_{ref}}r\left( t \right) = \\
 = {A_{ref}}{e_{ref}}\left( t \right) +\\
 +{B_i}\left[ {u\left( t \right) + \vartheta _i^{\rm{T}}\Psi \left( x \right) - K_i^xx\left( t \right) - K_i^rr\left( t \right)} \right] =\\
 ={A_{ref}}{e_{ref}}\left( t \right) + {B_i}\left[ {u\left( t \right) - \theta _i^{\rm{T}}\omega \left( t \right)} \right]{\rm{,}}
\end{array}
\end{equation}
where
\begin{displaymath}
\begin{array}{c}
{e_{ref}}\left( t \right) = x\left( t \right) - {x_{ref}}\left( t \right){\rm{,}}\\
\omega \left( t \right) = {{\begin{bmatrix}
{{x^{\rm{T}}}\left( t \right)}&{{r^{\rm{T}}}\left( t \right)}&{ - {\Psi ^{\rm{T}}}\left( x \right)}
\end{bmatrix}}^{\rm{T}}} \in {\mathbb{R}^{n + m + p}}{\rm{,}}\\
{\theta _i} = {{\begin{bmatrix}
{K_i^x}&{K_i^r}&{\vartheta _i^{\rm{T}}}
\end{bmatrix}}^{\rm{T}}} \in {\mathbb{R}^{\left( {n + m + p} \right) \times m}}{\rm{,}}
\end{array}    
\end{displaymath}
	
The equation \eqref{eq6} motivates to apply the control law with adjustable parameters:
\begin{equation}\label{eq7}
\begin{array}{c}
u\left( t \right) = {\hat \theta ^{\rm{T}}}\left( t \right)\omega \left( t \right){\rm{,}}
\end{array}
\end{equation}
where $\hat \theta \left( t \right) \in {\mathbb{R}^{\left( {n + m + p} \right) \times m}}$ is a continuous estimate of the piecewise constant parameters ${\theta _i}$.

The equation \eqref{eq7} is substituted into \eqref{eq6} to obtain:
\begin{equation}\label{eq8}
\begin{array}{c}
{\dot e_{ref}}\left( t \right) = {A_{ref}}{e_{ref}}\left( t \right) + B_{i}\left[ {{{\hat \theta }^{\rm{T}}}\left( t \right) - \theta _i^{\rm{T}}} \right]\omega \left( t \right) =\\
={A_{ref}}{e_{ref}}\left( t \right) + {B_i}\tilde \theta _i^{\rm{T}}\left( t \right)\omega \left( t \right){\rm{,}}
\end{array}    
\end{equation}
where ${\tilde \theta _i}\left( t \right) = \hat \theta \left( t \right) - {\theta _i}$ is the error of the parameter ${\theta _i}$ estimation.

The following assumption about the parameters ${\theta _i}$ and the regressor $\Phi \left( t \right)$ excitation is made to state the problem strictly.

\emph{ \textbf{Assumption 2.} Let $\exists {\overline \Delta _\theta } > 0,{\text{\;}}{T_{{\text{min}}}} > \mathop {\min }\limits_{\forall i \in \mathbb{N}} {T_i} > 0$ such that $\forall i \in \mathbb{N}$ simultaneously:}
\begin{enumerate}
    \item \emph{$t_{i + 1}^ +  - t_i^ +  \ge {T_{{\rm{min}}}}$, $\left\| {{\theta _i} - {\theta _{i - 1}}} \right\| = \left\| {\Delta _i^\theta } \right\| \le {\overline \Delta _\theta }{\rm{.}}$}
    \item \emph{$\Phi \left( t \right) \in {\rm{FE}}$ over $\left[ {t_i^ + {\rm{;\;}}t_i^ +  + {T_i}} \right]$ with excitation level ${\alpha _i}$.}
    \item \emph{$\Phi \left( t \right) \in {\rm{FE}}$ over $\left[ {\hat t_i^ + {\rm{;\;}}t_i^ +  + {T_i}} \right]$ with excitation level ${\overline \alpha _i}$, ${\alpha _i} > {\overline \alpha _i} > 0,{\rm{\;}}\hat t_i^ +  \in \left[ {t_i^ + {\rm{;\;}}t_i^ +  + {T_i}} \right)$.}
    \item \emph{$\cancel{\exists\;}t_i^{esc} \in \left[ {t_i^ + {\rm{;\;}}t_i^ +  + {T_i}} \right){\rm{\;}}\mathop {{\rm{lim}}}\limits_{t \to t_i^{esc}} \left\| {x\left( t \right)} \right\| = \infty $.}
\end{enumerate}

\emph{ \textbf{Goal.} It is required to ensure that the following inequality holds $\forall t \in \left[ {t_i^ +  + {T_i}{\rm{;\;}}t_{i + 1}^ + } \right)$ for the plant \eqref{eq3} under the conditions that Assumptions 1 and 2 are met:}
\begin{equation}\label{eq9}
\begin{array}{c}
\left\| {\xi \left( t \right)} \right\| \le {c_1}{e^{ - {c_2}\left( {t - t_i^ +  - {T_i}} \right)}}{\rm{,}}
\end{array}    
\end{equation}
\emph{where $\xi \left( t \right) = {{\begin{bmatrix}
{e_{ref}^{\rm{T}}\left( t \right)}&{ve{c^{\rm{T}}}\left( {{{\tilde \theta }_i}\left( t \right)} \right)}
\end{bmatrix}}^{\rm{T}}}$ is an augmented tracking error.}

\section{Main Result}
Extending the concept of exponentially stable adaptive control \cite{b15} to the class of systems with piecewise constant unknown parameters, it is required to generate a measurable linear regression equation with respect to the unknown parameters ${\theta _i}$ of the control law over time intervals $\left[ {\hat t_i^ + {\text{; }}t_{i + 1}^ + } \right]$ to achieve the goal \eqref{eq9}:
\begin{equation}\label{eq10}
\begin{array}{c}
\forall t \in \left[ {\hat t_i^ + {\text{;\;}}t_{i + 1}^ + } \right]{\text{\;}}\mathcal{Y}\left( t \right) = \Omega \left( t \right){\theta _i}{\text{,}}
\end{array}    
\end{equation}
where $\mathcal{Y} \left( t \right) \in {\mathbb{R}^{\left( {n + m + p} \right) \times m}}$ is a measurable regression output (regressand), $\Omega \left( t \right) \in \mathbb{R}$ is a measurable scalar regressor, such that $\forall t \in \left[ {t_i^ + {\text{ + }}{T_i}{\text{;\;}}\hat t_{i + 1}^ + } \right){\text{\;}}\Omega \left( t \right) \geqslant {\Omega _{LB}} > 0$.

To derive the equation \eqref{eq10}, first of all, the estimates of $\hat t_i^ + $ are required. The detection algorithm, which has been proposed in \cite{b20}, is applied to obtain them.

\emph{ \textbf{Proposition 1.} On the basis of the filter states:}
\begin{equation}\label{eq11}
\begin{array}{c}
\dot {\overline \Phi} \left( t \right) =  - l\overline \Phi \left( t \right) + \Phi \left( t \right){\text{,\;}}\overline \Phi \left( {\hat t_i^ + } \right) = {0_{m + n + p}}{\text{,\;}}l > 0,
\end{array}    
\end{equation}
\emph{normalized signals:}
\begin{equation}\label{eq12}
\begin{gathered}
  {{\overline \varphi }_n}\left( t \right) = {n_s}\left( t \right)\overline \varphi \left( t \right){\text{,\;}}\overline \varphi \left( t \right) = {{\begin{bmatrix}
  {{{\overline \Phi }^{\text{T}}}\left( t \right)}&{{e^{ - l\left( {t - \hat{t}_i^ + } \right)}}} 
\end{bmatrix}}^{\text{T}}}{\text{,\;}}\\
{n_s}\left( t \right) = \tfrac{1}{{1 + {{\overline \varphi }^{\text{T}}}\left( t \right)\overline \varphi \left( t \right)}}{\text{,}} \\ 
  {{\overline z}_n}\left( t \right) = {n_s}\left( t \right)\left[ {x\left( t \right) - l\overline x\left( t \right)} \right]{\text{,}}\\
  \overline x\left( t \right) = {\begin{bmatrix}
  {{I_{n \times n}}}&{{0_{n \times m}}}&{{0_{n \times p}}} 
\end{bmatrix}}\overline \Phi \left( t \right){\text{,}} \\ 
\end{gathered}
\end{equation}
\emph{and the dynamic regressor extension and mixing procedure:}
\begin{equation}\label{eq13}
\begin{gathered}
  z\left( t \right){\text{:}} = adj\left\{ {\omega \left( t \right)} \right\}\Upsilon \left( t \right){\text{,\;}}\Delta \left( t \right){\text{:}} = det \left\{ {\omega \left( t \right)} \right\}{\text{,}} \\ 
  \omega \left( t \right){\text{:}} = \int\limits_{\hat t_i^ + }^t {{e^{ - \sigma \left( {\tau  - \hat t_i^ + } \right)}}{{\overline \varphi }_n}\left( \tau  \right)\overline \varphi _n^{\text{T}}\left( \tau  \right)d\tau } {\text{,\;}}\\
  \Upsilon \left( t \right){\text{:}} = \int\limits_{\hat t_i^ + }^t {{e^{ - \sigma \left( {\tau  - \hat t_i^ + } \right)}}{{\overline \varphi }_n}\left( \tau  \right)\overline z_n^{\text{T}}\left( \tau  \right)d\tau }
\end{gathered}
\end{equation}
\emph{using the following indicator:}
\begin{equation}\label{eq14}
\epsilon\left( t \right) = \Delta \left( t \right){\overline \varphi _n}\left( t \right)\overline z_n^{\text{T}}\left( t \right) - {\overline \varphi _n}\left( t \right)\overline \varphi _n^{\text{T}}\left( t \right)z\left( t \right){\text{,}}
\end{equation}
\emph{the algorithm:}
\begin{equation}\label{eq15}
\begin{gathered}
  {\text{initialize:\;}}i \leftarrow 1,{\text{\;}}{t_{\operatorname{up}}} = \hat t_{i - 1}^ +  \\ 
  {\text{IF\;}}t - {t_{up}} \geqslant {\Delta _{pr}}{\text{\;AND\;}}\left\| {\epsilon\left( t \right)} \right\| > 0\\
  {\text{\;THEN\;}}\hat t_i^ + {\text{:}} = t + {\Delta _{pr}}{\text{,\;}}{t_{up}} \leftarrow t{\text{,\;}}i \leftarrow i + 1, \\ 
\end{gathered}
\end{equation}
\emph{ensure that $\forall i \in \mathbb{N}$ the following inequalities holds: $\hat t_i^ +  \geqslant t_i^ +$,  \linebreak $\tilde t_i^ +  = {\Delta _{pr}} \leqslant {T_i}{\text{}}$ if the regressor excitation is propagated $\overline \Phi \left( t \right) \in {\rm{FE}} \Rightarrow \overline \varphi \left( t \right) \in {\rm{FE}}$ and Assumption 2 is met.}

\emph{Proof of Proposition is postponed to Appendix.}

Having at hand the estimate $\hat t_i^ + $ and signals $z\left( t \right){\text{,\;}}\Delta \left( t \right)$, which are obtained using the filtering \eqref{eq13} with resetting at time instants $\hat t_i^ + $, the equation \eqref{eq10} can be derived.

\emph{ \textbf{Proposition 2.} Using \eqref{eq5}, on the basis of \eqref{eq13} \linebreak $\forall t \in \left[ {\hat t_i^ + {\text{;\;}}t_{i + 1}^ + } \right]$ the functions $\mathcal{Y}\left( t \right)$ and $\Omega \left( t \right)$ are defined as:}
\begin{displaymath}
\begin{gathered}
  \mathcal{Y}\left( t \right){\text{:}} = { {\begin{bmatrix}
  {adj\left\{ {z_B^{\text{T}}\left( t \right){z_B}\left( t \right)} \right\}z_B^{\text{T}}\left( t \right)\left( {\Delta \left( t \right){A_{ref}} - {z_A}\left( t \right)} \right)} \\ 
  {adj\left\{ {z_B^{\text{T}}\left( t \right){z_B}\left( t \right)} \right\}z_B^{\text{T}}\left( t \right)\Delta \left( t \right){B_{ref}}} \\ 
  {adj\left\{ {z_B^{\text{T}}\left( t \right){z_B}\left( t \right)} \right\}z_B^{\text{T}}\left( t \right){z_{B\vartheta }}\left( t \right)} 
\end{bmatrix}}^{\text{T}}}{\text{,}} 
\end{gathered}
\end{displaymath}
\begin{equation}
\begin{gathered}
  \Omega \left( t \right){\text{:}} = adj\left\{ {z_B^{\text{T}}\left( t \right){z_B}\left( t \right)} \right\}z_B^{\text{T}}\left( t \right)\Delta \left( t \right){B_i}=\\ 
  = adj\left\{ {z_B^{\text{T}}\left( t \right){z_B}\left( t \right)} \right\}z_B^{\text{T}}\left( t \right){z_B}\left( t \right){\text{ = }} \\ 
   = det\left\{ {z_B^{\text{T}}\left( t \right){z_B}\left( t \right)} \right\}. \\ 
\end{gathered} \label{eq16}
\end{equation}
\emph{with the following notation:}
\begin{equation}\label{eq17}
\begin{gathered}
  {z_A}\left( t \right) = {z^{\text{T}}}\left( t \right){\mathfrak{L}_A}{\text{,\;}}{z_B}\left( t \right) = {z^{\text{T}}}\left( t \right){\mathfrak{L}_B}{\text{,\;}}\\
  {z_{B{\vartheta ^{\text{T}}}}}\left( t \right) = {z^{\text{T}}}\left( t \right){\mathfrak{L}_{B\vartheta }}{\text{,}} \\ 
  {\mathfrak{L}_A} = {{\begin{bmatrix}
  {{I_{n \times n}}}&{{0_{n \times \left( {m + p + 1} \right)}}} 
\end{bmatrix}}^{\text{T}}}{\text{,\;}}\\
{\mathfrak{L}_B} = { {{{\begin{bmatrix}
  {{0_{m \times n}}}&{{1_{m \times m}}}&0_{m \times \left( {p + 1} \right)}\end{bmatrix}}}} ^{\text{T}}}{\text{,}} \\ 
  {\mathfrak{L}_{B\vartheta }} = {{\begin{bmatrix}
  {{0_{m \times \left( {n + m} \right)}}}&{{1_{m \times p}}}&{{0_{m \times 1}}} 
\end{bmatrix}}^{\text{T}}}{\text{,}} \\ 
\end{gathered} 
\end{equation}
\emph{and, if the implication $\overline \Phi \left( t \right) \in {\rm{FE}} \Rightarrow \overline \varphi \left( t \right) \in {\rm{FE}}{\text{}}$ \linebreak holds, then $\forall t \in \left[ {t_i^ + {\text{ + }}{T_i}{\text{;\;}}\hat t_{i + 1}^ + } \right){\text{\;}}{\Omega _{UB}} \geqslant \Omega \left( t \right) \geqslant {\Omega _{LB}} > 0.$}

\emph{Proof of Proposition 2 is presented in Appendix.}

Having obtained a regression equation with a scalar regressor $\Omega \left( t \right)$ that is a non-zero $\forall t \in \left[ {t_i^ + {\text{ + }}{T_i}{\text{;\;}}\hat t_{i + 1}^ + } \right)$, in accordance with the method of exponentially stable adaptive control \cite{b15, b17,b18,b19}, we are in position to introduce an adaptive law to calculate estimates $\hat \theta \left( t \right)$, which guarantees that the goal \eqref{eq9} is achieved.

\emph{ \textbf{Theorem 1.} Let Assumptions 2 is met and the regressor excitation is propagated $\overline \Phi \left( t \right) \in {\rm{FE}} \Rightarrow \overline \varphi \left( t \right) \in {\rm{FE}}$, then the adaptive law:}
\begin{equation}\label{eq18}
\begin{gathered}
  \dot {\hat \theta} \left( t \right) =  - \gamma \left( t \right)\Omega \left( t \right)\left( {\Omega \left( t \right)\hat \theta \left( t \right) - \Omega \left( t \right)\theta \left( t \right)} \right) =\\
  =- \gamma \left( t \right){\Omega ^2}\left( t \right)\tilde \theta \left( t \right){\text{,\;}} \\ 
  \gamma \left( t \right) = \left\{ \begin{gathered}
  {\text{0}}{\text{,\;if\;}}\Omega \left( t \right) \leqslant \rho {\text{,}} \hfill \\
  \frac{{{\gamma _0}{\lambda _{{\text{max}}}}\left( {\omega \left( t \right){\omega ^{\text{T}}}\left( t \right)} \right) + {\gamma _1}}}{{{\Omega ^2}\left( t \right)}}{\text{ otherwise}}{\text{,}} \hfill \\ 
\end{gathered}  \right. \\ 
\end{gathered}  
\end{equation}
\emph{when $\rho  \in \left( {0{\text{;\;}}{\Omega _{LB}}} \right]{\text{,\;}} {\gamma _0} \geqslant 1,{\text{\;\;}}{\gamma _1} \geqslant 0$ ensures the following properties $\forall i \in \mathbb{N}$:}
\begin{enumerate}
    \item ${t_a}{\text{,\;}}{t_b} \!\!\in \! \!\left[ {t_i^ +  + {T_i}{\text{;\;}}t_{i + 1}^ + } \right)\!\! \Rightarrow \!\! \forall {t_a}\!\! \geqslant\!\! {t_b}{\text{\;}}\left| {{{\tilde \theta }_k}\left( {{t_a}} \right)} \right|\!\!\! \leqslant \!\! \left| {{{\tilde \theta }_k}\left( {{t_b}} \right)} \right|{\text{,\;}}\linebreak k = \overline {1,\left( {n + m + p} \right)m} {\text{;}}$
    \item $\forall t \in \left[ {t_i^ + {\text{;\;}}t_i^ +  + {T_i}} \right]{\text{\;}}\left\| {\xi \left( t \right)} \right\| \leqslant {\xi _{UB}}{\text{;}}$
    \item $\forall t \in \left[ {t_i^ +  + {T_i}{\text{;\;}}t_{i + 1}^ + } \right){\text{\;}}\left\| {\xi \left( t \right)} \right\| \leqslant {c_1}{e^{ - {c_2}\left( {t - t_i^ +  - {T_i}} \right)}}.$
\end{enumerate}

\emph{Proof of Theorem is given in Appendix.}

Thus, the proposed adaptive control system consists of the measured signal $\Phi \left( t \right)$ processing equations \eqref{eq11}-\eqref{eq13}, detection algorithm \eqref{eq14}-\eqref{eq15}, equations \eqref{eq16} to calculate the regressor $\Omega \left( t \right)$ and regressand $\mathcal{Y}(t)$ of \eqref{eq10}, and adaptive law \eqref{eq18} to obtained the estimates $\hat{\theta}(t)$. The developed system: \linebreak 1) does not require knowledge of the switching time instants $t_i^ + $ as well as the control matrix $B$, and 2) ensures the required exponential boundedness \eqref{eq9} in case of finite excitation of the regressor somewhere inside the time interval between plant parameters switches.

\emph{ \textbf{Remark 1.} The effect of parameter $l$ value of the filter \eqref{eq11} on the regressor excitation propagation \linebreak $\overline \Phi \left( t \right) \in {\rm{FE}} \Rightarrow \overline \varphi \left( t \right) \in {\rm{FE}}$ was discussed in detail in \cite{b19}. The choice of arbitrary parameters $l{\text{,\;}}\sigma {\text{,\;}}{\Delta _{pr}}{\text{,\;}}\rho,\;{\gamma _0}{\text{,\;}}{\gamma _1}$ of the adaptive system should be made following the recommendations given in the related section of \cite{b20}.}

\emph{ \textbf{Remark 2.} The adaptive law \eqref{eq18} is BIBO stable and ensures boundedness of the error $\tilde \theta \left( t \right)$ in case the regression equation \eqref{eq10} is affected by bounded disturbances. Moreover, if $\tilde \theta \left( t \right)$ converges to a compact set that provides boundedness of $x\left( t \right)$, then \eqref{eq18} additionally guarantees interval boundedness of the error $\left\| {\xi \left( t \right)} \right\| \leqslant {c_1}{e^{ - {c_2}\left( {t - t_i^ +  - {T_i}} \right)}}{\text{ + }}{c_3}$.}

\emph{ \textbf{Remark 3.} If the number of parameter switches is finite $i \leqslant {i_{\max }}$, then, based on the results of Theorem, it is easy to show that $\forall t \geqslant t_{{i_{\max }}}^ +  + {T_{{i_{\max }}}}$ the law \eqref{eq18} guarantees global exponential stability $\left\| {\xi \left( t \right)} \right\| \leqslant {c_1}{e^{ - {c_2}\left( {t - t_{{i_{\max }}}^ +  - {T_{{i_{\max }}}}} \right)}}$.}

\section{Numerical Experiment}
A MIMO system with two switches of unknown parameters was considered ($x\left( 0 \right) = {{\begin{bmatrix}
  { - 1}&0 
\end{bmatrix}}^{\text{T}}}$):
\begin{equation}\label{eq19}
\dot x\left( t \right) = \left\{ \begin{gathered}
  {A_0}x\left( t \right) + {B_0}\left( {u\left( t \right) + \vartheta _0^{\text{T}}\Psi \left( x \right)} \right){\text{,\;}}t \in \left[ {{\text{0;\;5}}} \right) \hfill \\
  {A_1}x\left( t \right) + {B_1}\left( {u\left( t \right) + \vartheta _1^{\text{T}}\Psi \left( x \right)} \right){\text{,\;}}t \in \left[ {{\text{5;\;10}}} \right) \hfill \\
  {A_2}x\left( t \right) + {B_2}\left( {u\left( t \right) + \vartheta _2^{\text{T}}\Psi \left( x \right)} \right){\text{,\;}}t \geqslant 10 \hfill \\ 
\end{gathered}  \right.
\end{equation}
where the matrices of the unknown parameters, initial conditions and basis functions were defined as follows:
\begin{equation}\label{eq20}
\begin{gathered}
  {A_0} = {A_2} ={\begin{bmatrix}
  1&1 \\ 
  { - 1}&{ - 1} 
\end{bmatrix}}{\text{,\;}}{B_0} = {B_2} =  {\begin{bmatrix}
  {0.{\text{8}}}&{0.{\text{8}}} \\ 
  0&{0.{\text{8}}} 
\end{bmatrix}}{\text{,\;}}\\
{\vartheta _0} = {\vartheta _2} = {\begin{bmatrix}
  {0.2}&0 \\ 
  0&{ - 0.{\text{1}}} 
\end{bmatrix}}{\text{,\;}} {\vartheta _1} =  {\begin{bmatrix}
  { - 0.2}&0 \\ 
  0&{0.{\text{1}}} 
\end{bmatrix}},\\ 
  {A_1} = {\begin{bmatrix}
  { - 1}&{ - 1} \\ 
  1&1 
\end{bmatrix}}{\text{,\;}}{B_1} = {\begin{bmatrix}
  {0.{\text{8}}}&{ - 0.{\text{8}}} \\ 
  0&{ - 0.{\text{8}}} 
\end{bmatrix}}{\text{.\;}} \\ 
\end{gathered}
\end{equation}

The parameters of the reference model \eqref{eq4}, filters \eqref{eq11}, \eqref{eq13}, detection algorithm \eqref{eq15} and adaptive law \eqref{eq18} were set as:
\begin{equation}\label{eq21}
\begin{gathered}
  {A_{ref}}\! =\! {\begin{bmatrix}
  0&1 \\ 
  { - 4}&{ - 2} 
\end{bmatrix}}{\text{,\;}}{B_{ref}}\! =\! {\begin{bmatrix}
  4&0 \\ 
  0&4 
\end{bmatrix}}{\text{,\;}}r\left( t \right)\! = \! {\begin{bmatrix}
  1 \\ 
  {{e^{ - t}} - 1} 
\end{bmatrix}}{\text{,}} \\ 
  \hat \theta \left( 0 \right) = {\left[ {{0_{2 \times 2}}{\text{ }}{I_{2 \times 2}}{\text{\;}}{0_{2 \times 2}}} \right]^{\text{T}}}{\text{,\;}}l = 10,{\text{\;}}\sigma  = 5,{\text{\;}}\\
  {\Delta _{pr}} = 0.{\text{1}},\; \;\rho=10^{25}{\text{,\;}}{\gamma _0} = 1,{\text{\;}}{\gamma _1} = 1. \\ 
\end{gathered}
\end{equation}

Figure 1 presents the transients of $\hat t_i^ + $ and $\Omega \left( t \right)$.

   \begin{figure}[thpb]
      \centering
      \includegraphics[scale=0.95]{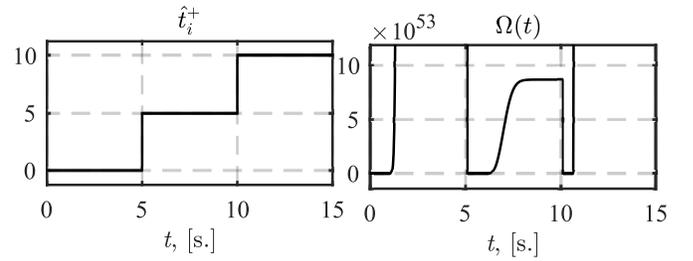}
      \caption{Transients of $\hat t_i^ + $ and $\Omega \left( t \right)$}
      \label{Figure1} 
      \end{figure}

The curves that are presented in Fig. 1 fully validated the conclusions made in Propositions 1 and 2.  The detection algorithm \eqref{eq15} was capable to detect switches of the system \eqref{eq3} parameters, and the regressor $\Omega \left( t \right)$ was nonzero after each such switch starting from some time instant if Assumption 2 was met.

Transients of $\left\| {ve{c^{\text{T}}}\left( {{{\tilde \theta }_i}\left( t \right)} \right)} \right\|$ and $\left\| {{e_{ref}}\left( t \right)} \right\|$ are shown in Figure 2.

   \begin{figure}[thpb]
      \centering
      \includegraphics[scale=0.95]{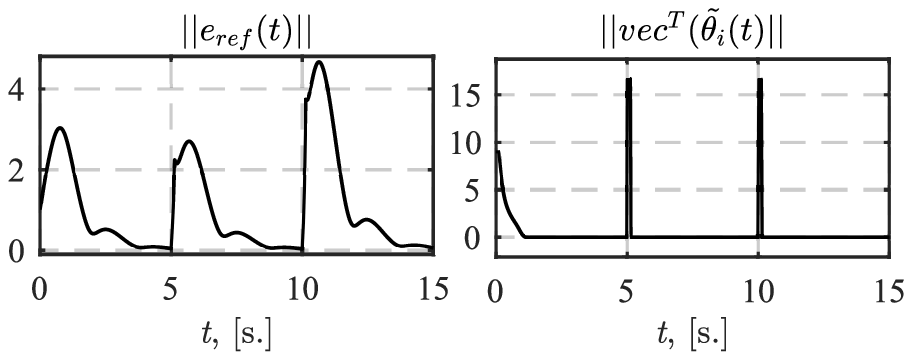}
      \caption{Transients of $\left\| {ve{c^{\text{T}}}\left( {{{\tilde \theta }_i}\left( t \right)} \right)} \right\|$ and $\left\| {{e_{ref}}\left( t \right)} \right\|$}
      \label{Figure2} 
      \end{figure}

The simulation results confirmed the Theorem. The augmented tracking error $\xi \left( t \right)$ did exponentially converge to zero over the time ranges between two consecutive changes of the unknown parameters.

\section{Conclusion}
The concept of the exponentially stable adaptive control \cite{b15} was extended to the class of MIMO switched systems with matched uncertainty and completely unknown control matrix. The solution was obtained by way of augmentation of the earlier-proposed parameterization with dynamic filters that reset their states at time instants, which were identified by a recently proposed algorithm to detect switching of the unknown parameters of the linear regression equations \cite{b20}. The proposed adaptive control system does not require the plant control matrix to be known, ensures the aperiodic transients of the adjustable parameters, and guarantees exponential convergence of the augmented tracking error over the time range between two consecutive switches in case the regressor is finitely exciting somewhere inside such interval. However, it can only be applied to systems without finite escape time over time range $\left[ {t_i^ + {\rm{;\;}}t_i^ +  + {T_i}} \right)$.

The scope of further research is to ensure the global exponential stability of the closed-loop adaptive control system instead of the interval exponential boundedness \eqref{eq9}.


\appendices

\renewcommand{\theequation}{A\arabic{equation}}
\setcounter{equation}{0}  

\section*{Appendix}
{\it Proof of Proposition 1.} In accordance with Lemma 6.8 in \cite{b1}, if $\Phi \left( t \right) \in {\rm{FE}}$, then $\overline \Phi \left( t \right) \in {\rm{FE}}$, and as, following the \linebreak statement of Proposition 1, $\overline \Phi \left( t \right) \in {\rm{FE}} \Rightarrow {\overline \varphi _n}\left( t \right) \in {\rm{FE}}$ \linebreak is also true, then $\Phi \left( t \right) \in {\rm{FE}} \Rightarrow {\overline \varphi _n}\left( t \right) \in {\rm{FE}}$. In \cite{b20} it has been proved that the implication ${\overline \varphi _n}\left( t \right) \in {\rm{FE}}\Rightarrow\Delta \left( t \right) \in {\rm{FE}}$ holds. According to Assumption 2, $\Phi \left( t \right) \in {\rm{FE}}$ over $\left[ {t_i^ + {\rm{;\;}}t_i^ +  + {T_i}} \right]{\rm{,\;}}\left[ {\hat t_i^ + {\rm{;\;}}t_i^ +  + {T_i}} \right]$, then ${\overline \varphi _n}\left( t \right) \in {\rm{FE}}{\rm{,\;}}\Delta \left( t \right) \in {\rm{FE}}$ over these time ranges.

As it was proved in \cite{b20}, the algorithm \eqref{eq15} $\forall i \in \mathbb{N}$ ensures that $\hat t_i^ +  \ge t_i^ +$, $ \tilde t_i^ +  = {\Delta _{pr}} \le {T_i}$ if ${\overline \varphi _n}\left( t \right) \in {\rm{FE}}{\rm{,\;}}\Delta \left( t \right) \in {\rm{FE}}$ over the time intervals $\left[ {t_i^ + {\rm{;\;}}t_i^ +  + {T_i}} \right]{\rm{,\;}}\left[ {\hat t_i^ + {\rm{;\;}}t_i^ +  + {T_i}} \right]$ and $\epsilon \left( t \right)$ is an indicator of the plant parameters switches.

Then, the only thing we need to prove is that $\epsilon\left( t \right)$ is such an indicator. To this end, the equation $\chi \left( t \right) = x\left( t \right) - l\overline x\left( t \right)$ is differentiated with respect to time:
\begin{equation}
\begin{array}{c}
\dot \chi \left( t \right) = \dot x\left( t \right) - l\dot {\overline x}\left( t \right) = \\={\Theta ^{\rm{T}}}\left( t \right)\Phi \left( t \right) + {l^2}\overline x\left( t \right) - lx\left( t \right) = \\
 =  - l\chi \left( t \right) + {\Theta ^{\rm{T}}}\left( t \right)\Phi \left( t \right){\rm{,\;}}\chi \left( t \right) = x\left( {\hat t_i^ + } \right)
\end{array}\normalsize{\label{eqA1}}
\end{equation}

Considering the time range $\left[ {t_i^ + {\rm{;\;}}\hat t_i^ + } \right)$, the solution of \eqref{eqA1} is substituted into the definition of ${\overline z_n}\left( t \right)\;\left( {i > 0,{\rm{ }}\forall t \in \left[ {t_i^ + {\rm{; }}\hat t_i^ + } \right)} \right)$:
\begin{equation}
\begin{array}{c}
{{\overline z}_n}\left( t \right) = {n_s}\left( t \right)\left( {{e^{ - l\left( {t - \hat t_{i - 1}^ + } \right)}}x\left( {\hat t_{i - 1}^ + } \right) + } \right.\\
 + \left. {\int\limits_{\hat t_{i - 1}^ + }^t {{\Theta ^{\rm{T}}}\left( t \right){e^{ - l\left( {t - \tau } \right)}}\Phi \left( \tau  \right)d\tau } } \right) = \\
 = {n_s}\left( t \right)\left( {{e^{ - l\left( {t - \hat t_{i - 1}^ + } \right)}}x\left( {\hat t_{i - 1}^ + } \right) + } \right.\\
 + \Theta _{i - 1}^{\rm{T}}\int\limits_{\hat t_{i - 1}^ + }^{t_i^ + } {{e^{ - l\left( {t - \tau } \right)}}\Phi \left( \tau  \right)d\tau }  + \\
\left. { + \Theta _i^{\rm{T}}\int\limits_{t_i^ + }^t {{e^{ - l\left( {t - \tau } \right)}}\Phi \left( \tau  \right)d\tau } } \right) = {{\overline \Theta }^{\rm{T}}}\left( t \right){{\overline \varphi }_n}\left( t \right) + \\
 + \underbrace {{n_s}\left( t \right)\left( {\Theta _{i - 1}^{\rm{T}} - \Theta _i^{\rm{T}}} \right)\int\limits_{\hat t_{i - 1}^ + }^{t_i^ + } {{e^{ - l\left( {t - \tau } \right)}}\Phi \left( \tau  \right)d\tau } }_{{{\overline \varepsilon }_0}\left( t \right)}.
\end{array}\normalsize{\label{eqA2}}
\end{equation}

The equation \eqref{eqA2} is substituted into definition of $z\left( t \right)$ to obtain $\left( {i > 0,{\rm{\;}}\forall t \in \left[ {t_i^ + {\rm{;\;}}\hat t_i^ + } \right)} \right)$:
\begin{equation}
\begin{array}{c}
{\rm{ }}z\left( t \right) = \Delta \left( t \right)\overline \Theta \left( t \right) + {{\overline \varepsilon }_1}\left( t \right){\rm{,\;}}{{\overline \varepsilon }_1}\left( t \right) = \\
 = adj\left\{ {\omega \left( t \right)} \right\}\left( {\int\limits_{\hat t_{i - 1}^ + }^t {{e^{ - \int\limits_{\hat t_{i - 1}^ + }^\tau  {\sigma ds} }}{{\overline \varphi }_n}\left( \tau  \right)\overline \varepsilon _0^{\rm{T}}\left( \tau  \right)} d\tau  + } \right.\\
 + \left. {\int\limits_{\hat t_{i - 1}^ + }^{t_i^ + } {{e^{ - \int\limits_{\hat t_{i - 1}^ + }^\tau  {\sigma ds} }}{{\overline \varphi }_n}\left( \tau  \right)\overline \varphi _n^{\rm{T}}\left( \tau  \right)} d\tau \left( {{{\overline \Theta }_{i - 1}} - {{\overline \Theta }_i}} \right)} \right).
\end{array}
\normalsize{\label{eqA3}}
\end{equation}

Finally, the equations \eqref{eqA2} and \eqref{eqA3} are substituted into \eqref{eq14} $\left( {i > 0,{\rm{\;}}\forall t \in \left[ {t_i^ + {\rm{;\;}}\hat t_i^ + } \right)} \right)$:
\begin{equation}
\epsilon\left( t \right) = \Delta \left( t \right){\overline \varphi _n}\left( t \right)\overline \varepsilon _0^{\rm{T}}\left( t \right) - {\overline \varphi _n}\left( t \right)\overline \varphi _n^{\rm{T}}\left( t \right){\overline \varepsilon _1}\left( t \right).
\normalsize{\label{eqA4}}
\end{equation}

By conducting similar to \eqref{eqA2}, \eqref{eqA3}, \eqref{eqA4} reasoning for the time range $\left[ {\hat t_i^ + {\rm{;\;}}t_{i + 1}^ + } \right)$, we have $ \overline \varepsilon _0^{\rm{T}}\left( t \right) = 0,{\rm{\;}}{\overline \varepsilon _1}\left( t \right) = 0$, and hence the interval-based definition of $\epsilon(t)$ is valid:
\begin{equation}
\forall i \in \mathbb{N}\;\epsilon\left( t \right){\rm{:}} = \left\{ \begin{array}{l}
\eqref{eqA4}{\rm{,\;}}i > 0,{\rm{\;}}\forall t \in \left[ {t_i^ + {\rm{;\;}}\hat t_i^ + } \right)\\
{\rm{0}}{\rm{,\;}}\forall t \in \left[ {\hat t_i^ + {\rm{;\;}}t_{i + 1}^ + } \right)
\end{array} \right.,
\normalsize{\label{eqA5}}
\end{equation}
from which it follows that the function $\epsilon\left( t \right)$ is indeed an indicator of the plant parameters switches (is nonzero only over the detection delay interval $\left[ {t_i^ + {\rm{; }}\hat t_i^ + } \right)$). So, according to the proof from \cite{b20} together with the fact that Assumption 2 is met, the inequalities $\forall i \in \mathbb{N} {\rm{\;}}\hat t_i^ +  \ge t_i^ + $, $\tilde t_i^ +  = {\Delta _{pr}} \le {T_i}$ hold when the algorithm \eqref{eq15} is applied.

{\it Proof of Proposition 2.} In accordance with the proof of Proposition 1, as far as the time range $\left[ {\hat t_i^ + {\rm{;\;}}t_{i + 1}^ + } \right)$  is concerned, we have ${\overline \varepsilon _1}\left( t \right) = 0$. Then the following definitions hold:
\begin{equation}
\begin{array}{c}
{z_A}\left( t \right) = {z^{\rm{T}}}\left( t \right){\mathfrak{L}_A} = \Delta \left( t \right){A_i}\\{z_B}\left( t \right) = {z^{\rm{T}}}\left( t \right){\mathfrak{L}_B} = \Delta \left( t \right){B_i}{\rm{,}}\\
{z_{B{\vartheta ^{\rm{T}}}}}\left( t \right) = {z^{\rm{T}}}\left( t \right){\mathfrak{L}}{_{B{\vartheta ^{\rm{T}}}}} = \Delta \left( t \right){B_i}\vartheta _{\mathop{\rm i}\nolimits} ^{\rm{T}} = {z_B}\left( t \right)\vartheta _{\mathop{\rm i}\nolimits} ^{\rm{T}}{\rm{,}}
\end{array}
\normalsize{\label{eqA6}}
\end{equation}
where ${z_A}\left( t \right) \in {\mathbb{R}^{n \times n}}{\rm{,\;}}{z_B}\left( t \right) \in {\mathbb{R}^{n \times m}}{\rm{,\;}}{z_{B{\vartheta ^{\rm{T}}}}}\left( t \right) \in {\mathbb{R}^{n \times p}}$ are measurable functions.

The third equation from \eqref{eqA6} is left-multiplied by $adj\left\{ {z_B^{\rm{T}}\left( t \right){z_B}\left( t \right)} \right\}z_B^{\rm{T}}\left( t \right)$, and each equation from \eqref{eq5} – by $adj\left\{ {z_B^{\rm{T}}\left( t \right){z_B}\left( t \right)} \right\}z_B^{\rm{T}}\left( t \right)\Delta \left( t \right)$. As a result, the definition \eqref{eq16} is obtained.

When $\Phi \left( t \right) \in {\rm{FE}}$ over $\left[ {t_i^ + {\rm{;\;}}t_i^ +  + {T_i}} \right]{\rm{,\;}}\left[ {\hat t_i^ + {\rm{;\;}}t_i^ +  + {T_i}} \right]$, then ${\overline \varphi _n}\left( t \right) \in {\rm{FE}}$, thus, if Assumption 2 is met, then according to \cite{b20} $\forall t \in \left[ {t_i^ + {\rm{ + }}{T_i}{\rm{;\;}}\hat t_{i + 1}^ + } \right){\rm{\;}}{\Delta _{UB}} \ge \Delta \left( t \right) \ge {\Delta _{LB}} > 0.$

The following equation holds for the regressor $\Omega \left( t \right)$:
\begin{equation}
\Omega \left( t \right) = det\left\{ {z_B^{\rm{T}}\left( t \right){z_B}\left( t \right)} \right\} = {\Delta ^n}\left( t \right)det\left\{ {B_i^{\rm{T}}{B_i}} \right\}.
\normalsize{\label{eqA7}}
\end{equation}

Owing to the controllability, we have $det\left\{ {B_i^{\rm{T}}{B_i}} \right\} \ne 0$. As a result, when $\Phi \left( t \right) \in {\rm{FE}}$ over $ \left[ {t_i^ + {\rm{;\;}}t_i^ +  + {T_i}} \right]{\rm{,\;}}\left[ {\hat t_i^ + {\rm{;\;}}t_i^ +  + {T_i}} \right]$ and excitation is propagated $\overline \Phi \left( t \right) \in {\rm{FE}} \Rightarrow {\overline \varphi _n}\left( t \right) \in {\rm{FE}}$, then $\forall t \in \left[ {t_i^ + {\rm{ + }}{T_i}{\rm{;\;}}\hat t_{i + 1}^ + } \right){\rm{\;}}{\Omega _{UB}} \ge \Omega \left( t \right) \ge {\Omega _{LB}} > 0$, which completes the proof.

{\it Proof of Theorem.} When $\rho  \in \left( {0{\rm{;\;}}{\Omega _{LB}}} \right]$, the differential equation with respect to $\tilde \theta_{i} \left( t \right)$ $ \forall t \in \left[ {t_i^ +  + {T_i}{\rm{;\;}}t_{i + 1}^ + } \right)$ is written as: $\dot {\tilde \theta}_{i}\left( t \right) =  - \left( {{\gamma _0}{\lambda _{{\rm{max}}}}\left( {\omega \left( t \right){\omega ^{\rm{T}}}\left( t \right)} \right) + {\gamma _1}} \right)\tilde \theta_{i} \left( t \right){\rm{,}}$ from which it immediately follows that the first statement of Theorem holds.

The time interval $\left[ {t_i^ + {\text{;\;}}t_i^ +  + {T_i}} \right)$ is divided into two ranges: $\left[ {t_i^ + {\text{;\;}}\hat t_i^ + } \right)$ and $\left[ {\hat t_i^ + {\text{;\;}}t_i^ +  + {T_i}} \right)$. Considering $\left[ {t_i^ + {\text{;\;}}\hat t_i^ + } \right)$, according to \eqref{eqA3}, the equation \eqref{eq13} is affected by the disturbance ${\bar \varepsilon _1}\left( t \right)$, which is bounded owing to the definition \eqref{eqA3} (more details can be found in the proof of Proposition 3 in \cite{b20}). Such perturbation influences the function $\mathcal{Y}\left( t \right)$ through the parametrization \eqref{eq16} in a multiplicative way:
\begin{equation}
\forall t \in \left[ {t_i^ + {\text{;\;}}\hat t_i^ + } \right){\text{\;}}\mathcal{Y}\left( t \right) = \Omega \left( t \right){\theta _i}{\text{ + }}d\left( t \right){\text{,}}
\normalsize{\label{eqA8}}
\end{equation}
where $d\left( t \right)$ is a new bounded disturbance.

Then the adaptive law \eqref{eq18} $\forall t \in \left[ {t_i^ + {\text{;\;}}\hat t_i^ + } \right)$ is written as:
\begin{equation}
{\dot {\tilde \theta} _i}\left( t \right) =  - \left( {{\gamma _0}{\lambda _{{\text{max}}}}\left( {\omega \left( t \right){\omega ^{\text{T}}}\left( t \right)} \right) + {\gamma _1}} \right)\left( {{{\tilde \theta }_i}\left( t \right) - \tfrac{{d\left( t \right)}}{{\Omega \left( t \right)}}} \right).
\small{\label{eqA9}}
\end{equation}

Therefore, it follows that $\forall t \in \left[ {t_i^ + {\text{;\;}}\hat t_i^ + } \right){\text{\;}}\left\| {{{\tilde \theta }_i}\left( t \right)} \right\| \leqslant \tfrac{{\left\| {d\left( t \right)} \right\|}}{{{\Omega _{LB}}}}$. 
Considering the time range $\left[ {t_i^ + {\text{;\;}}t_i^ +  + {T_i}} \right)$ and taking into account that $\forall t \in \left[ {t_i^ + {\text{;\;}}t_i^ +  + {T_i}} \right){\text{\;}}\Omega \left( t \right) < {\Omega _{LB}} \Rightarrow \dot {\hat \theta} \left( t \right) \equiv 0$, the conclusion is made that the parameter error is bounded $\forall t \in \left[ {t_i^ + {\text{;\;}}t_i^ +  + {T_i}} \right){\text{\;}}\left\| {\tilde \theta_{i} \left( t \right)} \right\| \leqslant \tfrac{{\left\| {d\left( t \right)} \right\|}}{{{\Omega _{LB}}}}$. According to the statement 4 of Assumption 2, the system \eqref{eq3} has no finite escape time over the time range $\left[ {t_i^ + {\text{;\;}}t_i^ +  + {T_i}} \right)$, so, considering the conservative case, the error ${e_{ref}}\left( t \right)$ tends to infinity with exponential rate only and bounded by its arbitrarily large but finite value at the right-hand bound of the time interval: $\left\| {{e_{ref}}\left( t \right)} \right\| \leqslant \left\| {{e_{ref}}\left( {t_i^ +  + {T_i}} \right)} \right\|$. So, then $\forall t \in \left[ {t_i^ + {\text{;\;}}t_i^ +  + {T_i}} \right]{\text{\;}}$ the inequality $\left\| {\xi \left( t \right)} \right\| \leqslant {\xi _{UB}}{\text{}}$ holds.

To prove the third statement of Theorem, the function is introduced over $\left[ {t_i^ +  + {T_i}{\rm{;\;}}t_{i + 1}^ + } \right)$:
\begin{equation}
\begin{array}{c}
V = e_{ref}^{\rm{T}}P{e_{ref}} + tr\left\{ {{{\tilde \theta_{i} }^{\rm{T}}}\tilde \theta_{i} } \right\}{\rm{,\;}}\\
\underbrace {{\lambda _{{\rm{min}}}}\left( H \right)}_{{\lambda _{\mathop{\rm m}\nolimits} }}{\left\| \xi  \right\|^2} \le V\left( {\left\| \xi  \right\|} \right) \le \underbrace {{\lambda _{{\rm{max}}}}\left( H \right)}_{{\lambda _M}}{\left\| \xi  \right\|^2}{\rm{,}} \\
H = {\rm{blockdiag}}\left\{ {P{\rm{,\;}}I} \right\}{\rm{,}}
\end{array}
{\label{eqA10}}
\end{equation}
where $P$ is from \eqref{eq2} with $D = 0.{\rm{5}}{I_{n \times n}}$, $B := B_{i},\; A := A_{ref}$.

Applying $tr\left( {AB} \right) = BA$, the derivative of \eqref{eqA10} with respect to \eqref{eq8}, \eqref{eq18} is written as:
\begin{equation} \small{\label{eqA11}}
\begin{array}{c}
\dot V = e_{ref}^{\rm{T}}\left( {A_{ref}^{\rm{T}}P + P{A_{ref}}} \right){e_{ref}} + 2e_{ref}^{\rm{T}}PB_{i}{{\tilde \theta_{i} }^{\rm{T}}}\omega  - \\
 - tr\left( {{{\tilde \theta_{i} }^{\rm{T}}}\gamma {\Omega ^2}\tilde \theta_{i} } \right) =  - \mu e_{ref}^{\rm{T}}P{e_{ref}} - \\
 - e_{ref}^{\rm{T}}Q{Q^{\rm{T}}}{e_{ref}} + tr\left( {2{{\tilde \theta_{i} }^{\rm{T}}}\omega e_{ref}^{\rm{T}}QK - {{\tilde \theta_{i} }^{\rm{T}}}\gamma {\Omega ^2}\tilde \theta_{i} } \right).
\end{array}
\end{equation} 
Then, completing the square in \eqref{eqA11}, it is obtained:
\begin{equation}\label{eqA12}
\small
\begin{array}{c}
\dot V =  - \mu e_{ref}^{\rm{T}}P{e_{ref}} + tr\left( { - {K^{\rm{T}}}{Q^{\rm{T}}}{e_{ref}}e_{ref}^{\rm{T}}QK + } \right.\\
 + 2{{\tilde \theta_{i} }^{\rm{T}}}\omega e_{ref}^{\rm{T}}QK\left. { \pm {{\tilde \theta_{i} }^{\rm{T}}}\omega {\omega ^{\rm{T}}}\tilde \theta_{i}  - {{\tilde \theta_{i} }^{\rm{T}}}\gamma {\Omega ^2}\tilde \theta_{i} } \right) = \\
 =  - \mu e_{ref}^{\rm{T}}P{e_{ref}} + tr\left( { - \left( {e_{ref}^{\rm{T}}QK - {{\tilde \theta_{i} }^{\rm{T}}}\omega } \right) \times } \right.\\
 \times {\left( {e_{ref}^{\rm{T}}QK - {{\tilde \theta_{i} }^{\rm{T}}}\omega } \right)^{\rm{T}}} + \left. {{{\tilde \theta_{i} }^{\rm{T}}}\omega {\omega ^{\rm{T}}}\tilde \theta_{i}  - {{\tilde \theta_{i} }^{\rm{T}}}\gamma {\Omega ^2}\tilde \theta_{i} } \right) \le \\
 \le  - \mu e_{ref}^{\rm{T}}P{e_{ref}} + tr\left\{ {{{\tilde \theta_{i} }^{\rm{T}}}\omega {\omega ^{\rm{T}}}\tilde \theta_{i}  - {{\tilde \theta_{i} }^{\rm{T}}}\gamma {\Omega ^2}\tilde \theta_{i} } \right\}\le \\
 \le  - \mu {\lambda _{\min }}\left( P \right){\left\| {{e_{ref}}} \right\|^2} - \left( {{\gamma _1} + \kappa } \right){\left\| {{{\tilde \theta }_i}} \right\|^2} \le  - \overline \eta V{\rm{,}}
\end{array}
\normalsize
\end{equation}
where ${\rm{0}} < \kappa  \le {\gamma _0}{\lambda _{\max }}\left( {\omega \left( t \right){\omega ^{\rm{T}}}\left( t \right)} \right) - \omega \left( t \right){\omega ^{\rm{T}}}\left( t \right)$, \linebreak $\eta  = \min\left\{ {{\textstyle{{\mu {\lambda _{\min }}\left( P \right)} \over {{\lambda _{\max }}\left( P \right)}}}{\rm{,\;}}{\gamma _1} + \kappa } \right\}$.

Therefore, it immediately follows that:
\begin{equation}
\left\| {\xi \left( t \right)} \right\| \le \sqrt {{\textstyle{{{\lambda _M}} \over {{\lambda _m}}}}} {e^{ - {\textstyle{\eta  \over 2}}\left( {t - t_i^ +  - {T_i}} \right)}}\left\| {\xi \left( {t_i^ +  + {T_i}} \right)} \right\|.
\small{\label{eqA12}}
\end{equation}
Having introduced the definition ${c_2} = {\textstyle{\eta  \over 2}},\;{c_1} =\linebreak= \sqrt {{\textstyle{{{\lambda _M}} \over {{\lambda _m}}}}} \left\| {\xi \left( {t_i^ +  + {T_i}} \right)} \right\|{\rm{,}}$ the equation \eqref{eq9} is obtained from \eqref{eqA12}, which completes the proof of Theorem.

\end{document}